\documentclass[twocolumn,showpacs,preprintnumbers,amsmath,amssymb]{revtex4}

\usepackage{graphicx}

\usepackage{dcolumn}  
\usepackage{bm}       

\newcommand{\be}{\begin{eqnarray}}
\newcommand{\ee}{\end{eqnarray}}
\newcommand{\ci}[1]{\cite{#1}}

\newcommand{\beq}{\begin{equation}}
\newcommand{\eeq}{\end{equation}}

\begin{document}

\title{Neutrino-nucleus interactions at low energies within Fermi-liquid theory}

\author{G.~I.~Lykasov}
\email{lykasov@jinr.ru}
\author{V.~A.~Bednyakov}
\email{bednyakov@jinr.ru}
\affiliation{JINR, Dubna, 141980, Moscow region, Russia}
\date{\today}

\begin{abstract}
   Cross sections are calculated for 
   neutrino scattering off heavy nuclei at energies below 50 MeV. 
   The theory of Fermi liquid is applied to estimate 
   the rate of neutrino-nucleon elastic and inelastic 
   scattering in a nuclear medium in terms of dynamic form factors. 
   The cross sections, obtained here in a rather simple way,
   are in agreement with 
   the results of the other much more sophisticated nuclear models. 
   A background rate from the solar neutrino interactions within 
   a large Ge detector is estimated in the above-mentioned approach.   
   The knowledge of the rate is in particular rather important 
   for new-generation large-scale neutrino experiments.   
\end{abstract}

\maketitle

\section{Introduction}
       Neutrino-nucleus reactions play an essential role 
       in various fields of particle, nuclear and neutrino physics
       like, for example, neutrino oscillation,
       beta  and double beta decay experiments, 
       search for the dark matter, investigation of the nucleon and 
       nuclear structure, etc.  
       A precise knowledge of the neutrino-nucleus cross sections 
       is necessary first of all 
       to correct interpretation of the results of above-mentioned 
       experiments.
       This is due to the fact that the neutrino-nucleus interactions
       in one experiment provide us 
       with a signal (when we, for example, look for neutrino 
       oscillations) and in another experiment constitute an
       inevitable background (when one, for example, looks for a dark matter
       signal).
       Therefore it is also obvious that the knowledge 
       of neutrino-nucleus cross sections 
       is very important
       when one studies the feasibility of a new neutrino project.
       
       At rather low neutrino energies, comparable to the nuclear
       excitation  energy, the neutrino-nucleus ($\nu A$) reactions
       are very sensitive to the nuclear structure and  the form of
       strong nucleon correlations in a nucleus. At these energies
       (several MeV)  the nuclear shell model is usually applied to
       calculate the  $\nu A$ cross sections (see for example
       \cite{walecka:72,klapdor:1986,langanke:04,caurier:05,volpe:04}).
       This model reproduces the Gamow-Teller  (GT) response for
       nuclei  \cite{caurier:99} and quite satisfactorily describes
       the available experimental data
       \cite{caurier:99,frekers:04,hagemann:04,baumer:03}.  At higher
       energies the Random Phase Approximation (RPA) is used to
       analyze the $\nu A$ processes  corresponding to the neutral and
       charged weak currents.  It describes the nuclear collective
       excitation 
       one-hole excitations of the correlated ground state.  This
       model and its modification, the continuum version CRPA,  are
       appropriate at neutrino energies above the nuclear excitation
       energy  \cite{kolbe:03}.  All these models are applied to
       investigate  various astrophysical aspects of nucleosynthesis,
       supernova collapse,  observation of solar and supernova
       neutrinos as well.  The most important ingredient of neutrino
       transport  calculations in such applications 
       the neutrino opacity at  supra-nuclear densities.
        
       Besides the above-mentioned nuclear models 
       there are a lot of models 
       also used to analyze neutrino propagation in dense matter 
       (see for example
\cite{sawyer,chris1,reddy:1998,reddy:1999,raffelt:95} and references therein). 
       They are very successfully applied to calculate the neutrino
       cross sections and neutrino mean free path 
       in dense infinite nuclear matter and neutron matter. 
       One of them is the model 
\cite{chris1} based on the Fermi liquid theory 
       suggested by L.D.Landau 
\cite{landau} and developed by A.B.Migdal 
\cite{migdal}, G.Baym and C.J.Pethick 
\cite{baym} and others. 
       The strong nucleon correlation  
       and the collective excitation mode are 
       included in this theory 
\cite{baym,chris:93}.

       In this paper we apply the Fermi liquid theory (FLT) 
       to calculate the cross sections of 
       neutrino scattering off heavy nuclei due  
       to the neutral and charged weak currents at rather 
       low neutrino energiers (below the Fermi energy). 
       In Sec.~II 
       we briefly present the general formalism for neutrino scattering in 
       dense matter in terms of dynamic form factors (FF). 
       In Sec.~III we calculate the vector and axial vector FF for symmetric
       nuclear matter within the FLT for neutral and charged current 
       neutrino processes.
       The procedure for the calculation of neutrino-nucleus cross sections 
       is described in Sec.~IV. 
       To illustrate the available application of our approach  
       we estimate the
       expected event rate 
       induced by the interaction of solar neutrinos with the $^{71}$Ge
       isotope 
       in a 1-kg target detector.
       In Sec.~V we compare our results with the results of 
       other different nuclear models. 
       The conclusion is presented in Sec.~VI.  
  
\section{General formalism} 
      The elastic or quasi-elastic neutrino-nucleon scattering is 
      due to the weak neutral current. 
      According to the Weinberg-Salam model, 
      the Lagrangian of such interaction has the form
\cite{chris1,weinb,salam,glashow}
\be
{\cal L}_l(x)=\frac{G_{\rm F}}{\sqrt{2}}\, l^\nu_\mu(x)j^\mu_z(x)~,
\label{def:wl}
\ee
      where $G_{\rm F}=1.166\times 10^{-5}$~GeV$^{-2}=1.436\times
      10^{-49}$~erg~cm$^3$ is the Fermi weak coupling constant,
\be
l^\nu_\mu(x)={\bar\psi}_\nu\gamma_\mu(1-\gamma_5)\psi_\nu
\label{def:lc}
\ee
is the lepton weak neutral current,
\be
j^\mu_z(x)=\frac{1}{2}\bar\psi_N\gamma^\mu(C_V-C_A\gamma_5)\psi_N
\label{def:nc}
\ee
    is the third component of the isospin current, 
    $C_V$, $C_A$ are the vector and axial coupling constants 
    for the neutral current reactions. 

    For the inelastic $\nu N$-scattering the Lagrangian is 
    related to the charged lepton and baryon weak currents
\be
{\cal L}_l(x)=\frac{G_{\rm F}}{\sqrt{2}}\, l_\mu(x)j^\mu_W(x)~,
\label{def:wl-ya}
\ee
     where
\be
j^\mu_W(x)=\bar\psi_N\gamma^\mu(_V-g_A\gamma_5)\psi_N~.
\label{def:nc-a}
\ee
     Here $g_V$, $g_A$ are the vector and axial-vector weak 
     coupling constants for the charged current reactions. 
     The lepton weak charged current has the following form
\be
l_\mu(x)={\bar\psi}_l\gamma_\mu(1-\gamma_5)\psi_\nu~.
\label{def:lc-a}
\ee
     At the low neutrino energies we may use the 
     following approximation for the neutral hadronic current
\ci{chris1}
\begin{equation}
\bar\psi_N\gamma^\mu(C_V-C_A\gamma_5)\psi_N
\rightarrow
C_V\psi_N^+\psi_N\delta_0^\mu-C_A\psi_N^+\sigma_i\psi_N\delta_i^\mu~.
\label{def:nrnc}
\end{equation}

    The similar approximation can also be used for the charge hadronic current. 
The rate of neutrino-nucleon scattering in a medium at low energies
can be presented in the following form (see, for example, 
\cite{chris1,raffelt:95,reddy:1999})
\be
W_{fi} & = & \frac{G_{\rm F}^2n}{4V}\left[ C_V^2(1+\cos\theta){\cal S}_V({\bf q},\omega)\right. \nonumber \\
& & \left. \qquad + C_{A}^2(3-\cos\theta){\cal S}_A({\bf q},\omega)\right]~,
\label{def:wfis}
\ee
    where $\theta$ is the scattering angle, 
    $V$ is the normalized volume, $n$ is the nuclear density,
    ${\cal S}_V$ and ${\cal S}_A$ are the vector and axial-vector 
    dynamic form factors (FF), 
    which depend on the transferred 
    3-momentum ${\bf q}$ and the energy transfer $\omega$.

    In principle, the hadronic vector current  
    ${\psi}^+_N\psi_N\delta_0^\mu$ 
    and the axial vector current  
    ${\psi}^+_N\sigma_i\psi_N\delta_i^\mu$ 
    can be modified in a medium. 
    The vector and axial coupling constants $C_V$ and $C_A$ 
    undergo renormalization;  
    however, their values can be changed in a medium 
    within 10--15\%  
\cite{arima1,sciavila1}.
    Therefore we neglect this effect. 
    We also neglect the antisymmetric part 
    ${\cal L}^a({\bf q},\omega)$ of the spin-spin dynamic form factor 
    because 
    it is proportional to ${\bf q}^2/m^2$
    \cite{chris1}
    and use only its symmetric part 
    ${\cal L}^s_{ij}({\bf q},\omega)
    \equiv\delta_{ij}{\cal S}_A({\bf q},\omega)$.

\section{Dynamic form factors in neutrino processes} 

\subsection{Neutral currents}
     
     First, we consider the FF for the neutral current neutrino processes, e.g., 
     the elastic and quasi-elastic neutrino-nucleon scattering in a medium.
     The FF for a one-component infinite nuclear system were
     constructed in Ref. \cite{chris1} to compute the neutrino mean free
     path in pure neutron matter for the densities of relevance for neutron 
     stars. In Ref. \cite{chris2} the FF were obtained  
     for two-component nuclear matter consisting of protons and neutrons also
     within the FLT. Within the RPA the FF in isospin-symmetric nuclear matter (SNM)
     were calculated in  Ref. \cite{reddy:1999}  by applying the Dyson type equation
     for the vector and axial vector polarization operators. Here we will demonstrate
     how the procedure suggested in Ref. \cite{reddy:1999} can be modified by applying the 
     FLT \cite{chris2} to get the FF for SNM.       

     The FF ${\cal S}_{V,A}$ are related to the corresponding response
     function $\chi^{}_{V,A}$  (see, for example, \cite{chris1,pines})
\beq
C_{V,A}^2{\cal S}_{V,A}(\omega,{\bf q})= 
                  \frac{2}{n }\,
		  \frac{{\rm Im}\chi^{}_{V,A}(\omega,{\bf q})}
		        {1-\exp(-\omega/T)}.
\label{def:ralschi}
\eeq  
    As was shown in Ref. \cite{reddy:1999}, the response
    functions $\chi^{}_V$ and $\chi^{}_A$ for the two-component 
    system consisting of protons and neutrons can be found as  the 
    solutions of the following Dyson perturbation equations for the
    spin-independent ${\cal F}$ and spin dependent ${\cal G}$ interactions
    of quasiparticles presented in the matrix form:
\begin{eqnarray}
\chi_V&=&\chi^0-\chi_V{\cal F}\chi^0,  \nonumber \\ 
\chi_A&=&\chi^0-\chi_A{\cal G}\chi^0, \label{def:chiva}
\end{eqnarray}
      Here $\chi^0$~=~ $\left(\begin{matrix}
                              \chi_p^0 & 0 \\
                              0 & \chi_n^0 
                              \end{matrix}\right)$ is the diagonal $2\times2$
      matrix consisting of $\chi_p^0$ and $\chi_n^0$ which are zero approximations 
      of the proton and neutron response functions in the interaction,  
      ${\cal F}$ and ${\cal G}$ are the spin-independent
      and spin-dependent interaction amplitudes. 
      For symmetric nuclear matter these amplitudes become the matrices
\cite{reddy:1999}:  
\beq
{\cal F}=\left(\begin{matrix}
               f_{pp} & f_{pn} \\
               f_{pn} & f_{nn} 
               \end{matrix}\right), \qquad  
\nonumber
{\cal G}=\left(\begin{matrix}
               g_{pp} & g_{pn} \\
               g_{pn} & g_{nn} 
               \end{matrix}\right),
\label{def:deq}
\eeq
     where $f_{pp}$, $f_{nn}$, $f_{pn}$ and 
     $g_{pp}$, $g_{nn}$, $g_{pn}$ are the spin-independent
     and spin-dependent 
     amplitudes of 
     $pp$, $nn$ and $pn$ interactions, respectively.
      
     Note that the amplitude of the interaction between 
     two quasi-particles $q$ and $q^\prime$ 
     with the three-momenta ${\bf p}$ and ${\bf p}^\prime$ 
     when the tensor forces are neglected has the following form  
\cite{landau,backman,migdal}:
\beq
f_{qq^\prime}({\bf p},{\bf p}^{\prime}) =
f+f^\prime({\bf\tau}\cdot{\bf\tau}^\prime)+
g({\bf\sigma}\cdot{\bf\sigma}^\prime)
+g^\prime({\bf\sigma}\cdot{\bf\sigma}^\prime)
({\bf\tau}\cdot{\bf\tau}^\prime)~,
\label{def:fg}
\eeq
   where   
   $f$, $f^\prime$, $g$, $g^\prime$ are the functions of only the angle $\theta$ between
   ${\bf p}$ and ${\bf p}^\prime$, 
   ${\bf\sigma},{\bf\sigma}^\prime$ and ${\bf\tau},{\bf\tau}^\prime$
   are the spin and isospin Pauli matrices. 
   The amplitudes associated with the proton-proton, neutron-neutron and 
   proton-neutron interactions have the following forms \cite{migdal}
\begin{eqnarray}
\nonumber
T_{pp}&=&T_{nn}=f_{pp}+g_{pp}\sigma\cdot\sigma^\prime,  \\.
T_{pn}&=&T_{np}=f_{pn}+g_{pn}\sigma\cdot\sigma^\prime~.
\label{def:fpp}
\end{eqnarray}
      From eqs. (\ref{def:fg},\ref{def:fpp}) we have \cite{migdal}
\begin{eqnarray}
\nonumber
     f_{pp}&=&f_{nn}=f+f^\prime, \\
     g_{pp}&=&g_{nn}=g+g^\prime, \\
\nonumber
     f_{pn}&=&f_{np}=f-f^\prime, \\
\nonumber
     g_{pn}&=&g_{np}=g-g^\prime~.  
\label{def:gpp}
\end{eqnarray}
      Inserting the matrices ${\cal F},{\cal G}$ and $\chi^0$ in
      (\ref{def:chiva}) we get a couple of equations for 
      $\chi_V^p\chi_V^n,$ and $\chi_A^p,\chi_A^n$.
\be
\chi_V^p(1+f_{nn}\chi_n^0)~+~\chi_V^nf_{pn}\chi_p^0~=~\chi_p^0~,\nonumber \\
\chi_V^pf_{pn}\chi_n^0~+~\chi_V^n(1+f_{pp}\chi_n^0)~=~\chi_n^0
\label{def:chivpn}
\ee
and
\be
\chi_A^p(1+g_{nn}\chi_n^0)~+~\chi_A^n g_{pn}\chi_p^0~=~\chi_p^0~,\nonumber \\
\chi_A^p g_{pn}\chi_n^0~+~\chi_A^n(1+g_{pp}\chi_n^0)~=~\chi_n^0~.
\label{def:chiapn}
\ee

      Formally solving equations (\ref{def:chivpn}) and (\ref{def:chiapn}) 
      (see 
\cite{reddy:1999,chris2} for details ) one gets the following general forms for 
      the vector and axial-vector response functions for symmetric nuclear matter
      $\chi_V^{s.n.m.}=\chi_V^p+\chi_V^n$ and $\chi_A^{s.n.m.}=\chi_A^p+\chi_A^n$ 
\begin{widetext}
\begin{eqnarray}
\chi_V^{s.n.m.} & = &\frac{(C_V^p)^2[1+f_{nn}\chi_n^0]\chi_p^0+(C_V^n)^2[1+f_{pp}\chi_n^0]\chi_n^0
-2C_V^pC_V^nf_{pn}\chi_n^0\chi_p^0}{1+f_{nn}\chi_n^0+f_{pp}\chi_p^0
+f_{pp}\chi_p^0f_{nn}\chi_n^0-f_{pn}\chi_n^0f_{pn}\chi_p^0}~,\label{def:chiV} \\ 
\chi_A^{s.n.m.} & = & \frac{(C_A^p)^2[1+g_{nn}\chi_n^0]\chi_p^0+(C_A^n)^2[1+g_{pp}\chi_p^0]\chi_n^0
-2C_A^pC_A^ng_{pn}\chi_n^0\chi_p^0}{1+g_{nn}\chi_n^0+g_{pp}\chi_p^0 
+g_{pp}\chi_p^0g_{nn}\chi_n^0-g_{pn}\chi_n^0g_{pn}\chi_p^0}~.
\label{def:chiA}
\end{eqnarray}
\end{widetext}
    Note that eq.(\ref{def:chiV}) and eq.(\ref{def:chiA}) for the response functions are similar in form to
    the equations for the polarization operators obtained in Ref.\cite{reddy:1999}.

    Actually, equations (\ref{def:chiV}) and (\ref{def:chiA}) can be solved within the Landau theory of the Fermi 
    liquid (LFLT) by relating the response function to the density fluctuation $\delta n^p$ and $\delta n^n$ for protons and 
    neutrons respectively and presenting (\ref{def:chiV}) and (\ref{def:chiA}) as the linearized quantum kinetic equations
    for these fluctuations. It was done in Ref.\cite{chris2} by expanding the quasiparticle interaction in Legendre 
    polynomials and keeping only the $l=0$ and $l=1$ terms. For symmetric nuclear matter $\chi_0^p=\chi_0^n\equiv \chi_0$ 
    calculated within the LFLT becomes \cite{chris1,chris2}
\begin{widetext}
\begin{eqnarray}
\chi_0 = N(0)g(\lambda)\equiv N(0)\frac{1}{2}\int_{-1}^1\frac{\mu d\mu}{\mu-\lambda} 
= N(0)\left[1-\frac{\lambda}{2}\ln\left|\frac{\lambda+1}{\lambda-1}\right|
+\frac{i\pi\lambda}{2}\theta(1-|\lambda|)\right],
\label{chipz}    
\end{eqnarray}
\end{widetext}
    where $\lambda=\omega/qv_F$ and $v_F=p_F/m^*$, $p_F$ is Fermi momentum of a quasiparticle, $m^*$ is the
    effective nucleon mass in a medium, $N(0)\equiv dn/d\epsilon_F$, $\epsilon_F=
   p_F^2/(2m^*)$ is the quasi-particle energy on the Fermi surface,
   $p_F$ depends on the density $n$.
   For SNM $N^s(0)=2 m^*p_F(n)/\pi^2$  and for the pure
   neutron matter (PNM) $N^p(0)= m^*p_F(n)/\pi^2$.

    Neglecting the terms of second order in the interaction of type $[f_{pp}f_{nn}-(f_{pn})^2]/3$
    and other similar terms \cite{chris2} one can get the following approximate forms for the response
    functions: 
\begin{widetext}
\begin{eqnarray}  
 \chi_V^{s.n.m.} & \simeq & \frac{N^s(0)}{V}\frac{C_V^2g(\lambda)+(C_V^-)^2(F\cdot g(\lambda))g(\lambda)+ 
(C_V^+)^2(F^\prime\cdot g(\lambda))g(\lambda)}{1+(F+F^\prime)\cdot g(\lambda)+
(F_0 g(\lambda))(F_0^\prime g(\lambda))}~,\label{def:chiv} \\
\chi_A^{s.n.m.} & \simeq & \frac{N^s(0)}{V}\frac{C_A^2g(\lambda)+ (C_A^-)^2(G\cdot g(\lambda))g(\lambda)+  
 (C_A^+)^2(G^\prime\cdot g(\lambda))g(\lambda)}{1+(G+G^\prime)\cdot g(\lambda)+
(G_0 g(\lambda))(G^\prime_0 g(\lambda))}~,
\label{def:chia} 
\end{eqnarray}
\end{widetext}
where $C_{V,A}^\pm=C_{V,A}^p\pm C_{V,A}^n, C_{V,A}^2=(C_{V.A}^p)^2+(C_{V,A}^n)^2$~.
\clearpage
Here
\begin{align}
(F\cdot g(\lambda))& =\left[ F_0+\frac{\lambda^2F_1}{1+(F_1+F_1^\prime)/3}\right]g(\lambda), \nonumber \\
(F^\prime\cdot g(\lambda))& =\left[ F_0^\prime+
\frac{\lambda^2F_1^\prime}{1+(F_1+F_1^\prime)/3}\right ]g(\lambda),\nonumber \\
(F+F^\prime)\cdot g(\lambda)& =\left[F_0+F_0^\prime+
\frac{\lambda^2(F_1+F_1^\prime)}{1+(F_1+F_1^\prime)/3}\right]g(\lambda)~,
\label{def:ffg}
\end{align}
   where $F_l=N^s(0)f_l,F^\prime_l=N^s(0)f^\prime_l,G_l=N^s(0)g_l,G^\prime_l=N^s(0)g^\prime_l (l=0,1)$ are 
   the Landau parameters; in fact, $f_l,f^\prime_l,g_l,g^\prime_l$ are the coefficients in the expansion of
   the interaction amplitude in Legendre polynomials.  
   Convolutions 
   $(G\cdot g((\lambda))$,
   $(G^\prime\cdot g((\lambda))$ and 
   $(G+G^\prime)\cdot g((\lambda)$
   have the same analytical forms as (\ref{def:ffg}), 
   but with $G_0,G _0^\prime,G_1,G_1^\prime$ instead of 
   $F_0,F _0^\prime,F_1,F_1^\prime$,
   respectively.
   More exact forms of these response functions are presented in
   Ref. \cite{chris2}.

   For free nucleons the weak coupling constants have 
   the following values:
   $C_V^p=0.08$, 
   $C_V^n=-1.0$, 
   $C_A^p=1.23$, $C_A^n=-1.23$. 
   In a medium they are renormalized within 
   10--20\% (see 
Ref. \cite{arima1}). 
   Therefore for simplicity we neglect this effect in our consideration. 
   Assuming further that 
   $C_A^2\simeq C_V^2$, we may present the axial response function in 
   the simple form
\begin{widetext}
\begin{eqnarray} 
\chi_A^{s.n.m.} & \simeq &  C_A^2\frac{N^s(0)}{V}\frac{g(\lambda)[1+4(G\cdot g(\lambda))]}
{1+(G+G^\prime)\cdot g(\lambda)+(G_0 g(\lambda)(G^\prime_0 g(\lambda))/2}~,
\label{def:chiaap}
\end{eqnarray}
\end{widetext}
    where $C_A=1.23$. 
    To get the response functions for pure neutron matter $\chi_V^{p.n.m.}$ 
    and $\chi_A^{p.n.m.}$ one can simply drop the proton-proton 
    interaction amplitudes; in practice, one should use 
    $f_{pp}=0,f_{pn}=0$ and
    $C_V^p=C_A^p=0$, $\chi_p^0=0$, $f^\prime=0,g^\prime=0$
    in 
(\ref{def:chiV}) and (\ref{def:chiA}).
    As a result one obtains
\cite{chris1}:
\begin{eqnarray}
\chi_V^{p.n.m.}&=&(C_V^n)^2\frac{N^p(0)}{V}\frac{g(\lambda)}{1+F\cdot g(\lambda)}
\label{def:chivn}, \\
\chi_A^{p.n.m.}&=&(C_A^n)^2\frac{N^p(0)}{V}\frac{g(\lambda)}{1+G\cdot 
g(\lambda)}~.
\label{def:chian}
\end{eqnarray}
       One can see that for nuclear matter the axial-vector
       response function 
       $\chi_A^{s.n.m.}$ is determined by the Landau parameters 
       $G_0,G_1,G^\prime_0,G^\prime_1$ linked to
       the spin and spin-isospin terms of the 
       interaction amplitude given by
eq.~(\ref{def:fg}).
       Therefore it is called the spin-isospin response function \cite{chris2}. 
       All necessary Landau parameters 
       for calculation of these response functions 
       can be taken from
\cite{backman,brown:2003} at the saturation density $n=2p_F^2/3\pi^2$.
       
      Let us discuss how the collective excitation in a 
      nuclear medium can be included. Within the LFLT the collective mode 
      excitation corresponds to the pole contribution 
      of the response function as a function of $\lambda$ at 
      $\lambda=\lambda_0\equiv c_s/v_F$, where $c_s$ is the zero sound velocity 
\cite{landau,baym,chris1,chris:93}. 
      If a pole occurs, the contribution
      of the collective mode is calculated as the residue of the response function 
      at  $\lambda=\lambda_0$
\cite{baym,chris1}. 
      One can show that at the saturation density $n=0.16$~fm$^{-3}$, 
      when $F_0+F_0^\prime < 0$ 
\cite{backman}, the effective field is attractive
      and the vector response function 
      $\chi_V^{s.n.m.}$ given by 
(\ref{def:chiv})  
      has no pole at real values of $\lambda$
\cite{baym,chris1,chris:93}. 
      Therefore, only single-pair excitations 
      exist for the density response spectrum at the normal density. 
      On the other hand, the spin-density or the axial-vector response 
      function $\chi_A^{s.n.m.}$ given by 
(\ref{def:chia}) has a pole at $\lambda=\lambda_0$ because
      $G_0+G_0^\prime > 0$
\cite{backman}. 
      We calculated 
      this spin-zero-sound pole contribution to the axial-vector
      FF as the residue of ${\cal S}_A$ at $\lambda=\lambda_0$. 
      The same procedure for inclusion of the collective excitation mode 
      in  pure neutron matter was proposed in 
\cite{chris1}.

\subsection{Charged currents} 
      Contrary to the elastic neutrino scattering the 
      isospin changing part of the particle-hole
      interaction is relevant for the charged current reactions. 
      In terms of the Fermi-liquid parameters,
      this means 
      that only $F^\prime$ and $G^\prime$ 
      contribute to the dynamic form factors \cite{reddy:1999}. 
      Therefore, to get the density response 
      and the spin-density functions one can use 
eq.~(\ref{def:chiv}) and 
eq.~(\ref{def:chia}) with 
      $F_0=0$, $G_0=0$ and with replacement of the coupling constants 
      $C_{V,A}^p$, $C_{V,A}^n$ by the weak coupling constants 
      corresponding to the charged current vector and axial couplings 
      $D_V$, $D_A$. The corresponding vector $\chi_V^{ch.,s.n.m}$
      and axial-vector $\chi_A^{ch.,s.n.m}$ response functions for symmetric 
      nuclear matter have the following forms:
\begin{widetext}
\begin{eqnarray}
\chi_V^{ch.,s.n.m.} & \simeq & \frac{N^s(0)}{V}
\frac{D_V^2g(\lambda)+(D_V^2g(\lambda)(F^c_0+\lambda^2 F^c_1/(1+F^c_1/3))g(\lambda)}
     {1+(F^c_0+\lambda^2 F^c_1/(1+F^c_1/3))g(\lambda)}~, \label{def:chivch} \\
\chi_A^{ch.,s.n.m.} & \simeq & \frac{N^s(0)}{V}
\frac{D_Ag(\lambda)+(D_A^2g(\lambda)(G^c_0+\lambda^2 G^c_1/(1+G^c_1/3))g(\lambda)}
     {1+ (G^c_0+\lambda^2 G^c_1/(1+G^c_1/3))g(\lambda)}~. \label{def:chiach} 
\end{eqnarray}
\end{widetext}
      Here $D_V=1.,D_A=1.23$ and $F^c_{0,1}=2 F^\prime_{0,1},G^c_{0,1}=2 G^\prime_{0,1}$ where 
      the factor 2 arises due to isospin consideration \cite{reddy:1999}. In 
      eqs.(\ref{def:chivch},\ref{def:chiach}) the terms of the second order of interaction of type
      $(F^c_{0,1})^2/3$ and $(G^c_{0,1})^2/3$ are also neglected as it was assumed 
      in eqs.(\ref{def:chiv},\ref{def:chia}).

      The rate of the neutrino-nucleon scattering in a medium due to the charged currents will have the 
      same form as the rate for neutral current processes given by eq.(\ref{def:wfis}) but multiplied
      by a factor 4 and with ${\cal S}_{V,A}$ replaced by ${\cal S}^{ch}_{V,A}$ related to
      $\chi_{V,A}^{ch.,s.n.m.}$. The factor 4 is due to the difference between the neutral and charged
      weak baryon currents (see eqs.(\ref{def:nc},\ref{def:nc-a}). 
            
\section{Neutrino cross section} 
     Using the forms obtained for the FF
     one can calculate the neutrino mean free path $l$ 
     in the nuclear matter. 
     According to Ref.\cite{chris1}, $1/l$ 
     is related to the neutrino rate $W_{fi}$:
\begin{eqnarray}
1/l=V\int\frac{d^3q}{(2\pi)^3}W_{fi}~.
\label{def:mnfp}
\end{eqnarray}
     With this quantity one can estimate the cross 
     section of the neutrino interaction with a heavy nucleus
     $\sigma_{\nu A}$.
\begin{eqnarray}
\sigma_{\nu A}=\frac{V_A}{l}~=~V_A\int\frac{d^3q}{(2\pi)^3}{\tilde W}_{fi}~,
\label{def:sig}
\end{eqnarray}
    where ${\tilde W}_{fi}=V_A\cdot W_{fi}$ and $V_A=A \cdot v_N$. 
    Here $A$ is the number of nucleons in a nucleus and 
    $v_N=4\pi/3r_N^3$ is the nucleon volume, 
    $r_N$ is the nucleon radius about 0.8~fm. 
    To calculate the elastic $\nu-A$ cross section we substitute into
    eq.(\ref{def:sig}) the neutrino rate ${\tilde W}_{fi}$ due 
    to the neutral weak and baryon currents, whereas for the absorption $\nu-A$ cross
    section ${\tilde W}_{fi}$ is the neutrino rate due to the charge
    exchange currents corresponding to the absorption processes of type
    $\nu~+~n\rightarrow e^-~+~p$. We calculated the total neutrino cross section 
    as the sum of the elastic and absorption $\nu-A$ cross sections. 
    Note that calculating the integral (\ref{def:sig}) we included the kinematics
    for the elastic and inelastic neutrino scattering off a nucleus.

    We calculated ${\cal S}_V$ and  ${\cal S}_A$ using 
    their relation to the corresponding response functions ${\chi}_V$ and
    ${\chi}_A$ given by eq.(\ref{def:ralschi}) and putting $T=0$. If the nucleus is 
    isotopically symmetric then the response functions can be calculated using the
    approximate formulas (\ref {def:chiv},\ref{def:chia}) and (\ref {def:chivch},\ref{def:chiach})
    or the exact expressions 
    presented in Ref. \cite{chris2}. Note that, as follows from our calculations the quantitative
    difference between these two forms for ${\chi}_{V,A}$ is negligibly small.
    In this paper we computed the cross sections of neutrino interactions with slightly
    asymmetric nuclei $^{71}Ge$ and $^{40}Ar$. To take into account the isotopic asymmetry 
    for these nuclei, we introduced fractions of protons and neutrons in front of $p-p$,
    $n-n$ and $p-n$ interactions solving the couple of equations (\ref{def:chiva}). The final
    forms for ${\chi}_{V,A}$ are slightly changed in comparison to eqs.(\ref{def:chiv},\ref{def:chia})
    when these fractions are included. We took the Landau parameters $F_0=-0.4,F_0^\prime=0.2,G_0=0.19,
    G_0^\prime=0.78, F_1=-0.66$ and $F^\prime_1=0.14$ from Ref. \cite{backman} and $F^\prime_1=0.14$
    from Ref. \cite{brown:2003} calculated for symmetric nuclear matter at the saturation nuclear density
    $n=2p_F^3/(3\pi^2)=0.16 fm^{-3}$. Note that $F_0,F_0^\prime,G_0,G_0^\prime$ are related to the compressibility,
    symmetry energy \cite{blaizot,sjoberg:1973} and spin susceptibility \cite{fantoni:03} respectively,
    whereas $F_1=3(m^*/m-1)$ \cite{baym,pines}. The parameters $G_1=0.48,G_1^\prime=0.08$ were taken from
    Ref. \cite{brown:2003}, using the relation of $G_0,G^\prime_0$ to $G_1,G^\prime_1$. Unfortunately, there 
    are no calculations of the Landau parameters for asymmetric nuclear matter. The isotopic asymmetry for
    $^{71}Ge$ and $^{40}Ar$ is small, therefore we assume that one can use the above results for the
    parameters in question obtained for symmetric nuclear matter.      
    
    To estimate the number of neutrino interactions $R$ per second within a detector we use the formula
\begin{eqnarray}
 R~=~
M_T N_A\sigma^{tot}_{\nu A}f_\nu~.
\label{def:R}
\end{eqnarray}
       Here $f_\nu$ denotes the initial neutrino flux presented in Fig.3,
       $N_A$ is the Avogadro constant, $M_T$ is the detector mass,
       $\sigma^{tot}_{\nu A}$ is the total cross section of neutrino scattering
       off a heavy nuclei A. 

\section{Results and discussions}
\begin{figure*}[t]
  \begin{tabular}{cc}
      \mbox{\includegraphics[angle=270,width=0.47\linewidth]{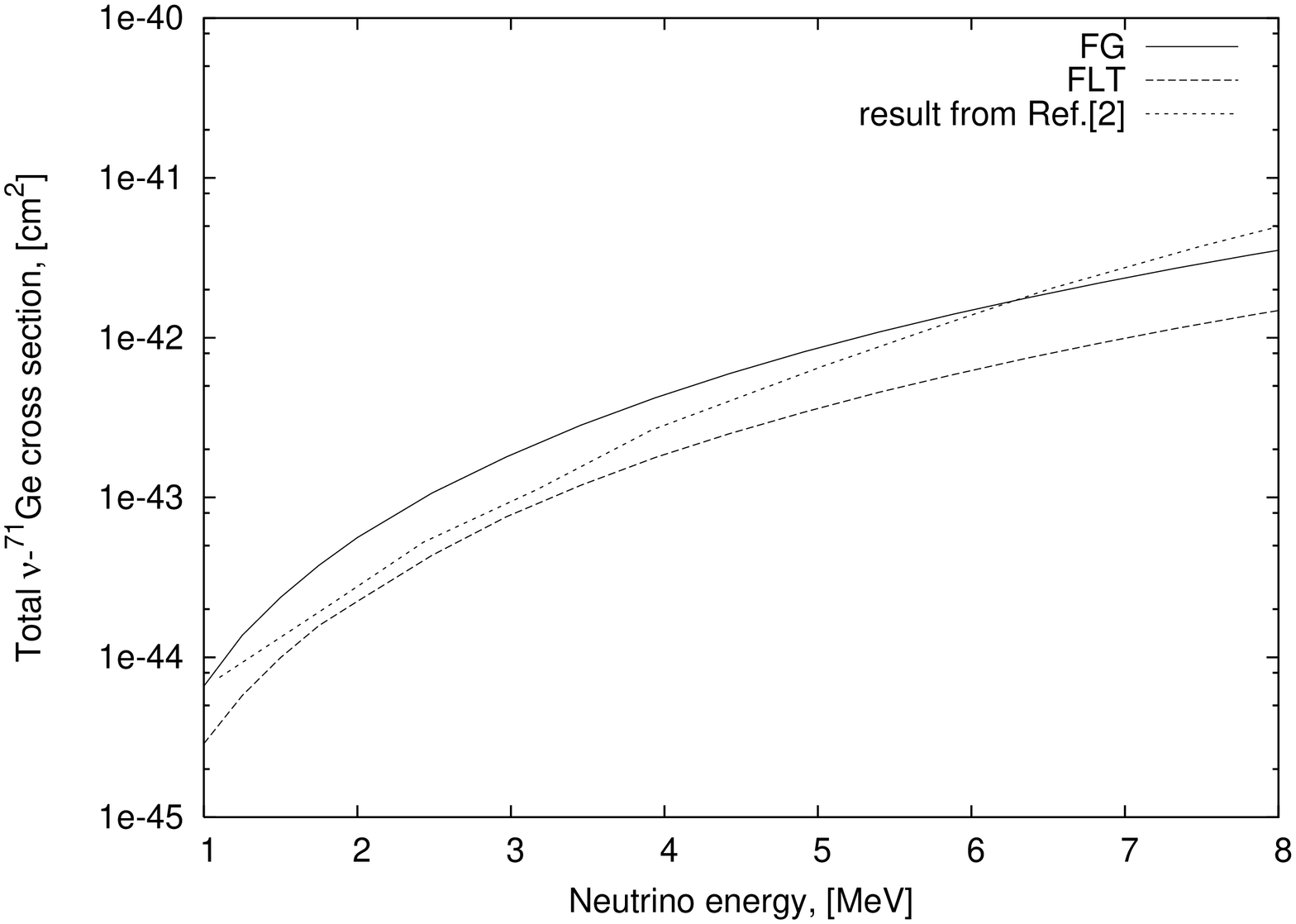}} &
      \mbox{\includegraphics[angle=270,width=0.47\linewidth]{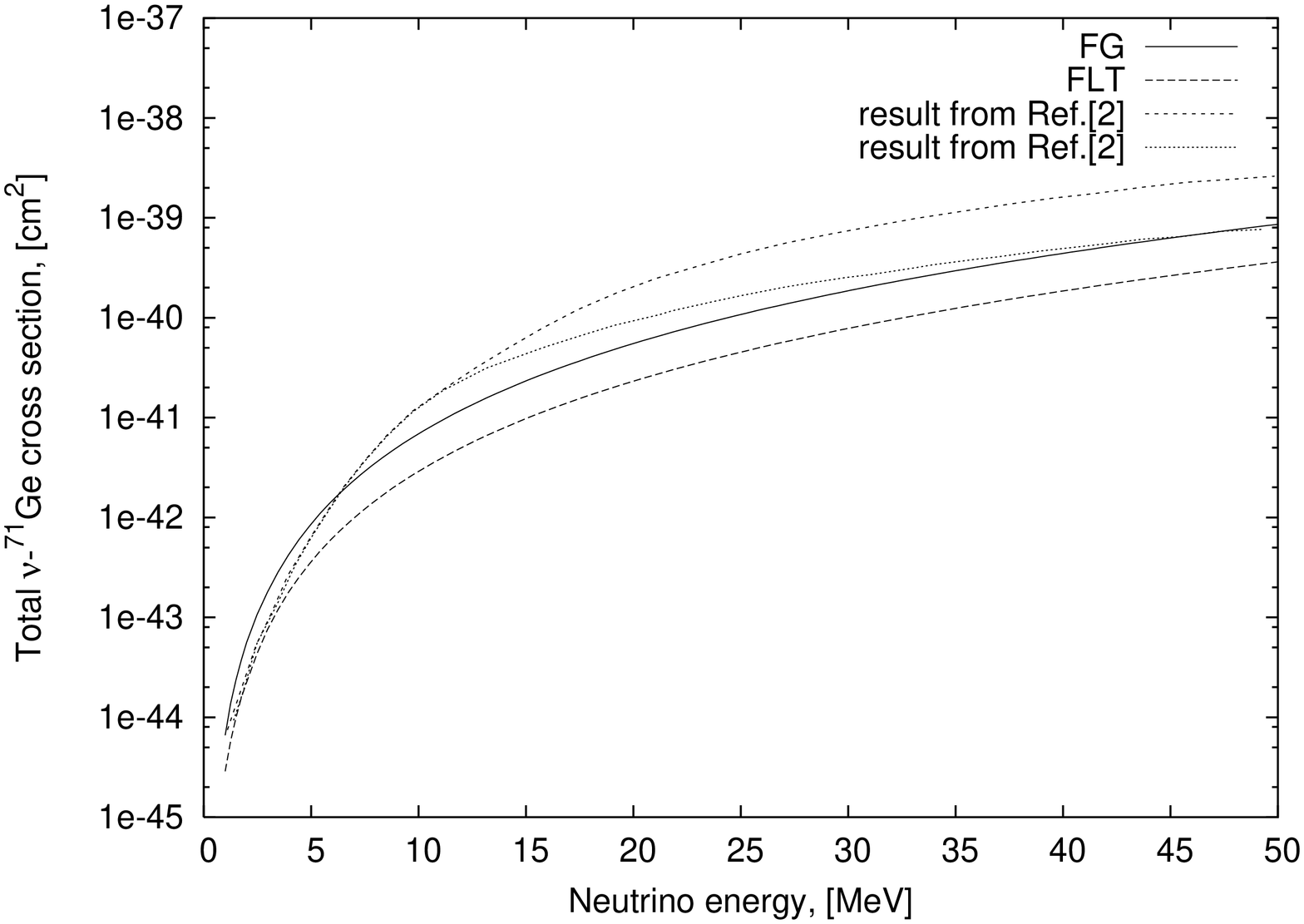}} \\
  \end{tabular}
\caption{\label{fig:fig1}
The total $\nu$-$^{71}$Ge cross section as a function of the neutrino energy $E_\nu$.
The solid line corresponds to the Fermi-gas approximation (FG), the long-dashed curve 
is the calculation within the Fermi liquid theory (FLT), the short-dashed line is the 
calculation within the shell model including the short-range part of the $N N$ 
interaction which is due to the $\pi$- and $\rho$-meson exchange \cite{klapdor:1986}, 
whereas the dotted line is the results of \cite{klapdor:1986} without these isobaric effects.
}
\vspace*{0.5cm}
  \begin{tabular}{cc}
      \mbox{\includegraphics[angle=270,width=0.47\linewidth]{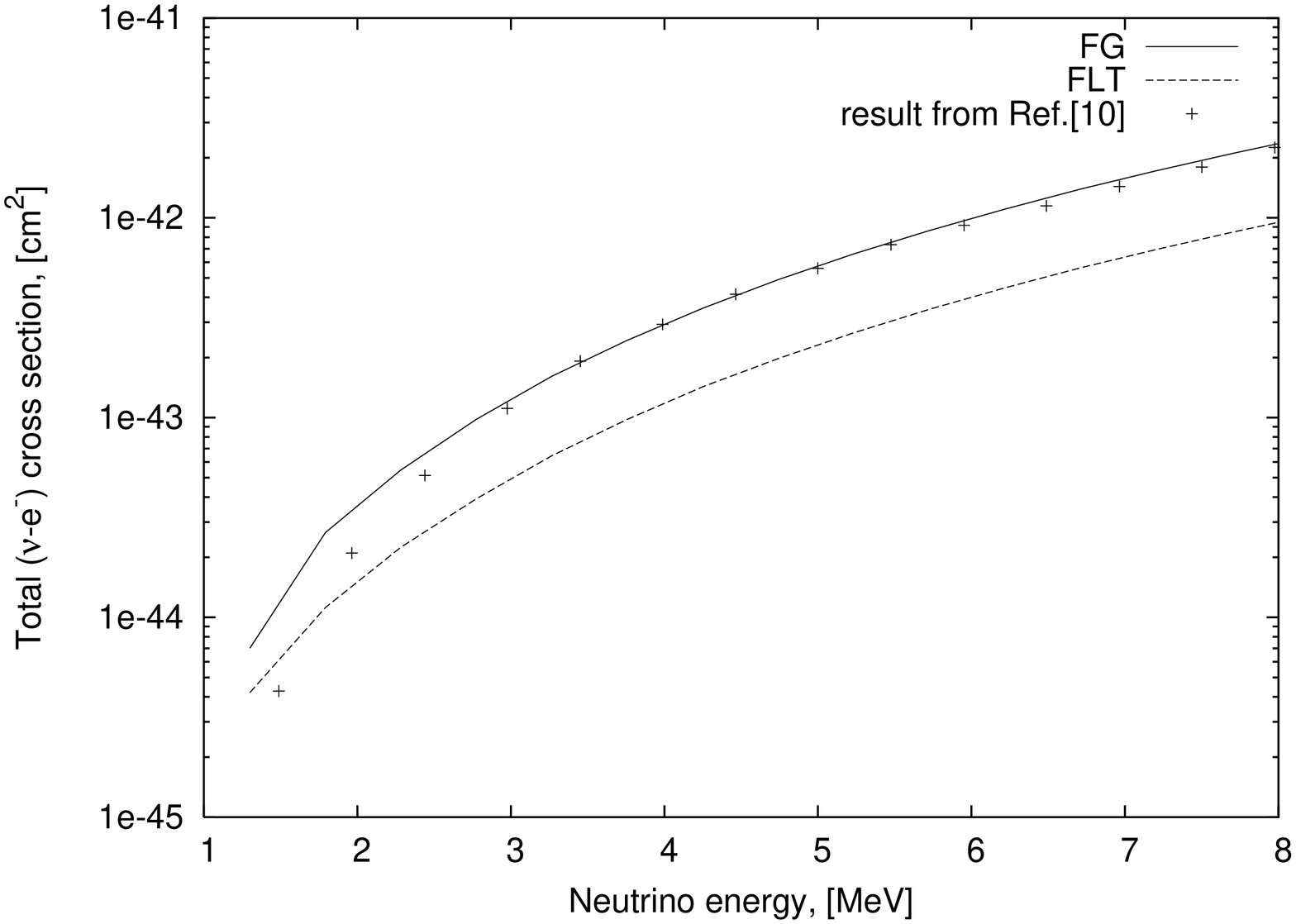}} &
      \mbox{\includegraphics[angle=270,width=0.47\linewidth]{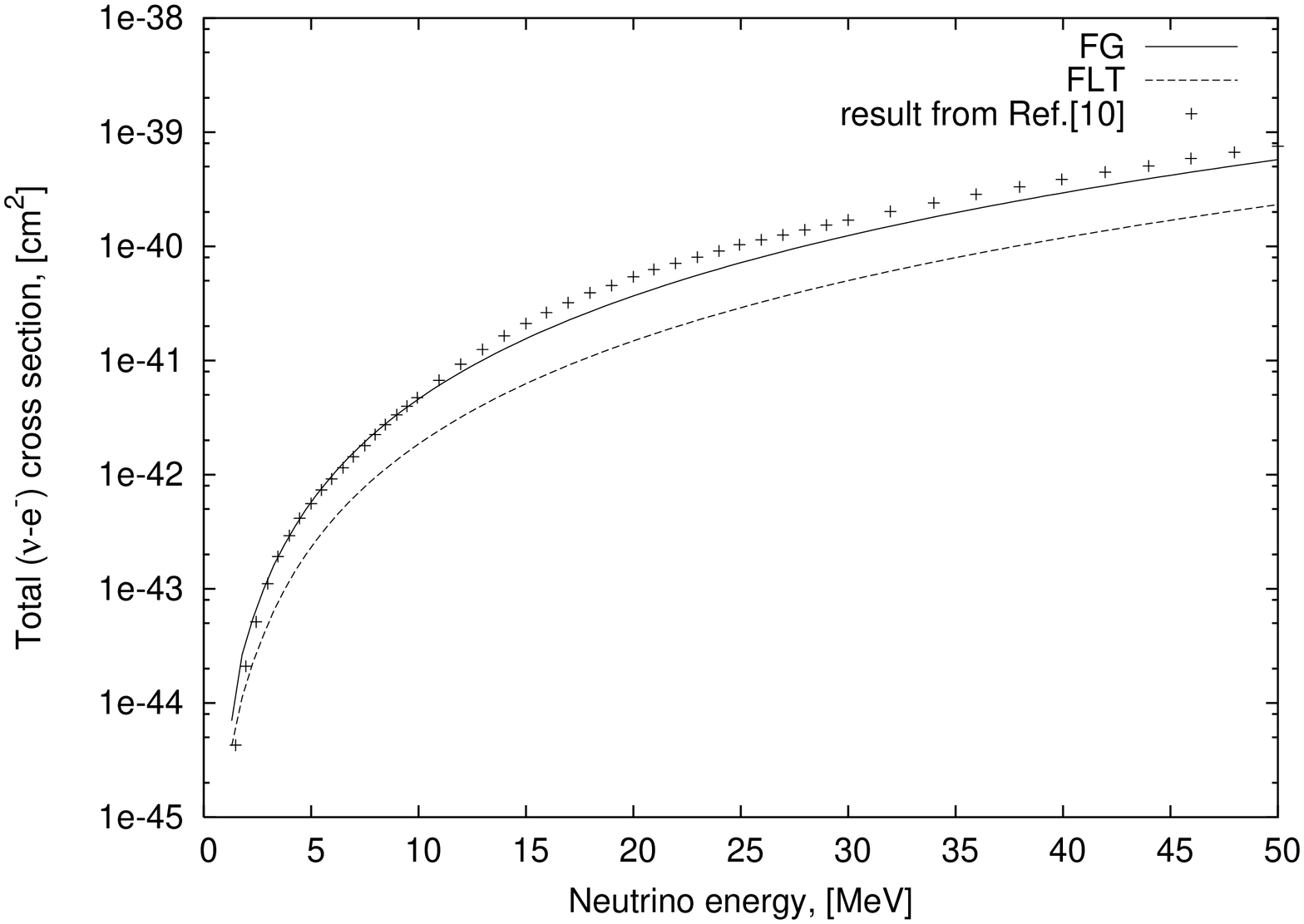}} \\
  \end{tabular}
\caption{\label{fig:fig2}
The total absorption $\nu$-$^{40}$Ar cross section as a function of the neutrino 
energy $E_\nu$.The solid line is the FG approximation, the long-dashed curve is the 
FLT calculation, the crosses correspond to the calculation of \cite{kolbe:03}
within the RPA.
}
\end{figure*}

    In Fig.~\ref{fig:fig1} the total $\nu$-$^{71}$Ge cross section including both elastic and
    absorption parts is presented
    as a function of the neutrino energy $E_\nu$ .
    The solid line corresponds to the Fermi gas approximation (FG), 
    e.g., when the nucleon-nucleon correlations in the $^{71}$Ge nucleus
    are neglected.
    The long-dashed curve is the calculation within the framework of the FLT
    when the tensor forces and the multi-pair contributions  are neglected.
    The inclusion of the two-particle-hole spin- and spin-isospin 
    correlations can increase the cross section.   
\cite{chris2}.
    Therefore the realistic total $\nu$-$^{71}$Ge 
    cross section calculated within the FLT
    including all these effects can differ from the Fermi gas 
    approximation by a factor of 1.5--2.
    The short-dashed line in Fig.~1 is the result of 
\cite{klapdor:1986}
    obtained within the shell model including the short-range part of $NN$ interaction,
    which is due to the $\pi$- and $\rho$-meson exchange, 
    and the possible isobar excitation,
    whereas the dotted curve (right panel of Fig.1) is the result of
\cite{klapdor:1986} without this isobaric effects.
    Note that at the low neutrino energies the calculation of 
    possible excitation of $\Delta$- or other $N^*$-isobars inside 
    a nucleus has some uncertainty related to 
    big off-mass-shell effects. 
    Therefore, the result corresponding to the short-dashed curve in
Fig.~1 (right panel) can be overestimated at $E_\nu > 10 MeV$. 
    One can see from the left panel of 
Fig.~1 that for solar neutrinos at energies less than 5 MeV 
    the FLT results in the total cross section values close 
    to the shell model calculation presented in 
\cite{klapdor:1986}. 
    At larger neutrino energies our  calculation
    within the FG approximation (the solid line in Fig.~1)
    is similar to the result of 
Ref. \cite{klapdor:1986} without the isobar excitation effect 
    (the dotted line in right panel of
Fig.~1).
    
    In
Fig.~2 the total absorption $(\nu,e^-)$cross section on $^{40}$Ar is presented. 
    The solid line corresponds to the FG approximation, 
    whereas the long-dashed curve is the calculation within the FLT. 
    Crosses correspond to the calculation of \cite{kolbe:03} within the RPA.     
    One can see from the left panel of 
Fig.~2 that at $E_\nu~<~ 3$~MeV the result from
\cite{kolbe:03} is similar to our calculations within the FLT, 
    whereas at larger energies 
(see right panel of Fig.~2) it does not contradict
    our results obtained within the FG approximation.    

    Figure~3  
(see, for example, 
\cite{Bahcall:2004fg}) 
    illustrates the solar neutrino flux continuum as a 
    function of the neutrino energy $E_\nu$.

    In
    Fig.~4 the total number of $^8$B-$\nu$ interactions $R$ per  month and per
    MeV in 1 kg of the $^{71}Ge$ detector is presented as a function of the 
    neutrino energy $E_\nu$. 
    The estimation of $R$ was done using eq.(\ref{def:R}). 
    The solid curve corresponds to the calculations when 
    the total neutrino cross section is calculated 
    within the FG approximation, whereas the
    long-dashed line is the FLT calculation of $\sigma_{\nu A}$. 
    The short-dashed curve corresponds to 
    $\sigma^{tot}_{\nu A}$ taken from \cite{klapdor:1986}. 
    A small difference between the FLT calculations and
    the results of 
\cite{klapdor:1986} is similar to the one for the cross sections presented
    in the left panel of 
Fig.~1. 
    For instance, with the results presented in Fig.~4 one can estimate 
    the total number of neutrino events expected (from $^8$B-$\nu$ flux)
    in the energy region 1--8 MeV
    within 1 Tonn (100 kg) of $^{71}$Ge detector 
    during one month. 
    If we also include $pep$-neutrinos (see Fig.3), the total number of neutrino
    events is about 20--30 (2--3) events per month, or 240--360 (24--36) per year.
    These values do not look negligibly small. 
    Note that the main contribution to these numbers comes from the energy region 
    $E_\nu\simeq 8 MeV$, where the FG reproduces correctly the cross sections 
    obtained within other models \cite{klapdor:1986,kolbe:03} (see the Figs.(1,2)). 
    It is due to the fact that $R$ falls down very fast when the neutrino energy 
    decreases. Therefore the discussed estimation of this neutrino-induced background
    in the energy interval from 1 MeV to 8 MeV is mainly determined by the FG calculation and
    is not sensitive to the calculations done within the FLT. On the other hand, 
    at lower neutrino energies the FLT gives a significantly smaller neutrino background 
    (by a factor of 2) than the FG calculation. It can be seen from Fig.1 (left panel).      
    The discussed estimation is rather approximate, nevertheless it illustrates
    the importance of the precise knowledge of solar 
    neutrino induced background for very accurate neutrino experiments.        
 
\section{Conclusion}
    In this paper we applied the Fermi liquid theory (FLT) 
    to calculate the cross sections of
    neutrino scattering off heavy nuclei at energies less 
    than 50 MeV. 
    Our results are in agreement with 
    the other more sophisticated calculations based on 
    the nuclear shell model and the RPA.
    The difference between our results and the other calculations
    is rather small in the neutrino energy $E_\nu ~<~$~5~MeV. 
    Nevertheless, our approach allows calculations of 
    the cross sections in the most simple way 
    comparing with the other nuclear models. 
    At 5~$<E_\nu<$~50~MeV the FLT results 
    differ from the RPA \cite{kolbe:03} and the simple shell model 
    (with neglection of the isobar excitations) \cite{klapdor:1986}
    by a factor of 2, whereas the use of the Fermi gas approximation 
    (FG) gives the difference much less than a factor 2. 
    It allows us to obtain a quite reliable estimate of the
    background induced by solar neutrinos with $E_\nu~<$~8--10~MeV
    in, for example, a Ge or Ga detector.
    
    The knowledge of such a background is very important for  very
    accurate neutrino experiments like, for example, 
    double $\beta$-decay experiments, 
    search for dark matter, etc. 
    One can conclude that for solar neutrinos the 
    FG approximation leads to underestimation of the 
    cross section by a factor of 2, 
    whereas at higher energies the results obtained within 
    the FG are closer to the simple shell model  \cite{klapdor:1986} and RPA
    calculations \cite{kolbe:03}.
    Therefore, the FLT can be safely used to estimate the background from solar
    neutrinos in heavy-nucleus detectors. 

    In this paper we estimated for the first time
    $\nu$-$A$ cross sections in the FLT, which can be used for present and future
    neutrino experiments. 
    We showed that this approach did not contradict to other
    nuclear models. 
    Unfortunately, there are no direct experimental data 
    on total cross sections of solar neutrino interactions
    with heavy nuclei at $E_\nu <$~10 MeV. 
    The existing data apply to the inverse $\beta$-decay 
    on a nucleus integrated over the energy spectrum of the neutrino produced 
    in $\beta$-decay of heavy nuclei, for example $^{238}$U.
    Therefore, in the future we plan to perform 
    a comparison of $\nu$-$A$ cross sections obtained 
    within the FLT and other nuclear models with available experimental data. 
    We are also going to improve our 
    approach by including possible
    multi-pair contributions to the weak response 
    and to calculate collective excitations 
    in a nucleus more carefully.              

\smallskip
\begin{acknowledgements}
 We are grateful to C.J.~Pethick, K.~Langanke 
 for very useful discussions and V.V.~Lyubushkin for his helps.
 V.B. acknowledges the support of the RFBR (grant RFBR-06-02-04003).
\end{acknowledgements}

\bibliography{paper_lb}

\begin{thebibliography}{32}
\expandafter\ifx\csname natexlab\endcsname\relax\def\natexlab#1{#1}\fi
\expandafter\ifx\csname bibnamefont\endcsname\relax
  \def\bibnamefont#1{#1}\fi
\expandafter\ifx\csname bibfnamefont\endcsname\relax
  \def\bibfnamefont#1{#1}\fi
\expandafter\ifx\csname citenamefont\endcsname\relax
  \def\citenamefont#1{#1}\fi
\expandafter\ifx\csname url\endcsname\relax
  \def\url#1{\texttt{#1}}\fi
\expandafter\ifx\csname urlprefix\endcsname\relax\def\urlprefix{URL }\fi
\providecommand{\bibinfo}[2]{#2}
\providecommand{\eprint}[2][]{\url{#2}}

\bibitem[{\citenamefont{Walecka}(1972)}]{walecka:72}
\bibinfo{author}{\bibfnamefont{J.~D.} \bibnamefont{Walecka}},
  \emph{\bibinfo{title}{Muon Physics}} (\bibinfo{publisher}{Academic Press},
  \bibinfo{address}{New York}, \bibinfo{year}{1972}).

\bibitem[{\citenamefont{Grotz et~al.}(1986)\citenamefont{Grotz, Klapdor, and
  Metzinger}}]{klapdor:1986}
\bibinfo{author}{\bibfnamefont{K.}~\bibnamefont{Grotz}},
  \bibinfo{author}{\bibfnamefont{H.~V.} \bibnamefont{Klapdor}},
  \bibnamefont{and}
  \bibinfo{author}{\bibfnamefont{J.}~\bibnamefont{Metzinger}},
  \bibinfo{journal}{Phys. Rev.} \textbf{\bibinfo{volume}{C33}},
  \bibinfo{pages}{1263} (\bibinfo{year}{1986}).

\bibitem[{\citenamefont{Langanke and Martinez-Pinedo}(2004)}]{langanke:04}
\bibinfo{author}{\bibfnamefont{K.}~\bibnamefont{Langanke}} \bibnamefont{and}
  \bibinfo{author}{\bibfnamefont{G.}~\bibnamefont{Martinez-Pinedo}},
  \bibinfo{journal}{Nucl. Phys.} \textbf{\bibinfo{volume}{A731}},
  \bibinfo{pages}{365} (\bibinfo{year}{2004}).

\bibitem[{\citenamefont{Caurier et~al.}(2005)\citenamefont{Caurier,
  Martinez-Pinedo, Nowacki, Poves, and Zuker}}]{caurier:05}
\bibinfo{author}{\bibfnamefont{E.}~\bibnamefont{Caurier}},
  \bibinfo{author}{\bibfnamefont{G.}~\bibnamefont{Martinez-Pinedo}},
  \bibinfo{author}{\bibfnamefont{F.}~\bibnamefont{Nowacki}},
  \bibinfo{author}{\bibfnamefont{A.}~\bibnamefont{Poves}}, \bibnamefont{and}
  \bibinfo{author}{\bibfnamefont{A.~P.} \bibnamefont{Zuker}},
  \bibinfo{journal}{Rev. Mod. Phys.} \textbf{\bibinfo{volume}{77}},
  \bibinfo{pages}{427} (\bibinfo{year}{2005}).

\bibitem[{\citenamefont{Serreau and Volpe}(2004)}]{volpe:04}
\bibinfo{author}{\bibfnamefont{J.}~\bibnamefont{Serreau}} \bibnamefont{and}
  \bibinfo{author}{\bibfnamefont{C.}~\bibnamefont{Volpe}},
  \bibinfo{journal}{Phys. Rev.} \textbf{\bibinfo{volume}{C70}},
  \bibinfo{pages}{055502} (\bibinfo{year}{2004}).

\bibitem[{\citenamefont{Caurier et~al.}(1999)\citenamefont{Caurier, Langanke,
  Martinez-Pinedo, and Nowacki}}]{caurier:99}
\bibinfo{author}{\bibfnamefont{E.}~\bibnamefont{Caurier}},
  \bibinfo{author}{\bibfnamefont{K.}~\bibnamefont{Langanke}},
  \bibinfo{author}{\bibfnamefont{G.}~\bibnamefont{Martinez-Pinedo}},
  \bibnamefont{and} \bibinfo{author}{\bibfnamefont{F.}~\bibnamefont{Nowacki}},
  \bibinfo{journal}{Nucl. Phys.} \textbf{\bibinfo{volume}{A653}},
  \bibinfo{pages}{439} (\bibinfo{year}{1999}).

\bibitem[{\citenamefont{Grewe et~al.}(2004)}]{frekers:04}
\bibinfo{author}{\bibfnamefont{E.~W.} \bibnamefont{Grewe}}
  \bibnamefont{et~al.}, \bibinfo{journal}{Phys. Rev.}
  \textbf{\bibinfo{volume}{C69}}, \bibinfo{pages}{064325}
  (\bibinfo{year}{2004}).

\bibitem[{\citenamefont{Hagemann et~al.}(2005)}]{hagemann:04}
\bibinfo{author}{\bibfnamefont{M.}~\bibnamefont{Hagemann}}
  \bibnamefont{et~al.}, \bibinfo{journal}{Phys. Rev.}
  \textbf{\bibinfo{volume}{C71}}, \bibinfo{pages}{014606}
  (\bibinfo{year}{2005}).

\bibitem[{\citenamefont{Baumer et~al.}(2003)}]{baumer:03}
\bibinfo{author}{\bibfnamefont{C.}~\bibnamefont{Baumer}} \bibnamefont{et~al.},
  \bibinfo{journal}{Phys. Rev.} \textbf{\bibinfo{volume}{C68}},
  \bibinfo{pages}{031303} (\bibinfo{year}{2003}).

\bibitem[{\citenamefont{Kolbe et~al.}(2003)\citenamefont{Kolbe, Langanke,
  Martinez-Pinedo, and Vogel}}]{kolbe:03}
\bibinfo{author}{\bibfnamefont{E.}~\bibnamefont{Kolbe}},
  \bibinfo{author}{\bibfnamefont{K.}~\bibnamefont{Langanke}},
  \bibinfo{author}{\bibfnamefont{G.}~\bibnamefont{Martinez-Pinedo}},
  \bibnamefont{and} \bibinfo{author}{\bibfnamefont{P.}~\bibnamefont{Vogel}},
  \bibinfo{journal}{J. Phys.} \textbf{\bibinfo{volume}{G29}},
  \bibinfo{pages}{2569} (\bibinfo{year}{2003}).

\bibitem[{\citenamefont{Sawyer}(1989)}]{sawyer}
\bibinfo{author}{\bibfnamefont{R.~F.} \bibnamefont{Sawyer}},
  \bibinfo{journal}{Phys. Rev.} \textbf{\bibinfo{volume}{C40}},
  \bibinfo{pages}{865} (\bibinfo{year}{1989}).

\bibitem[{\citenamefont{Iwamoto and Pethick}(1982)}]{chris1}
\bibinfo{author}{\bibfnamefont{N.}~\bibnamefont{Iwamoto}} \bibnamefont{and}
  \bibinfo{author}{\bibfnamefont{C.~J.} \bibnamefont{Pethick}},
  \bibinfo{journal}{Phys. Rev.} \textbf{\bibinfo{volume}{D25}},
  \bibinfo{pages}{313} (\bibinfo{year}{1982}).

\bibitem[{\citenamefont{Reddy et~al.}(1998)\citenamefont{Reddy, Prakash, and
  Lattimer}}]{reddy:1998}
\bibinfo{author}{\bibfnamefont{S.}~\bibnamefont{Reddy}},
  \bibinfo{author}{\bibfnamefont{M.}~\bibnamefont{Prakash}}, \bibnamefont{and}
  \bibinfo{author}{\bibfnamefont{J.~M.} \bibnamefont{Lattimer}},
  \bibinfo{journal}{Phys. Rev.} \textbf{\bibinfo{volume}{D58}},
  \bibinfo{pages}{013009} (\bibinfo{year}{1998}).

\bibitem[{\citenamefont{Reddy et~al.}(1999)\citenamefont{Reddy, Prakash,
  Lattimer, and Pons}}]{reddy:1999}
\bibinfo{author}{\bibfnamefont{S.}~\bibnamefont{Reddy}},
  \bibinfo{author}{\bibfnamefont{M.}~\bibnamefont{Prakash}},
  \bibinfo{author}{\bibfnamefont{J.~M.} \bibnamefont{Lattimer}},
  \bibnamefont{and} \bibinfo{author}{\bibfnamefont{J.~A.} \bibnamefont{Pons}},
  \bibinfo{journal}{Phys. Rev.} \textbf{\bibinfo{volume}{C59}},
  \bibinfo{pages}{2888} (\bibinfo{year}{1999}).

\bibitem[{\citenamefont{Raffelt and Seckel}(1995)}]{raffelt:95}
\bibinfo{author}{\bibfnamefont{G.}~\bibnamefont{Raffelt}} \bibnamefont{and}
  \bibinfo{author}{\bibfnamefont{D.}~\bibnamefont{Seckel}},
  \bibinfo{journal}{Phys. Rev.} \textbf{\bibinfo{volume}{D52}},
  \bibinfo{pages}{1780} (\bibinfo{year}{1995}).

\bibitem[{\citenamefont{Landau}(1957)}]{landau}
\bibinfo{author}{\bibfnamefont{L.}~\bibnamefont{Landau}}, \bibinfo{journal}{Zh.
  Eksp. Teor. Fiz.} \textbf{\bibinfo{volume}{32}}, \bibinfo{pages}{101}
  (\bibinfo{year}{1957}).

\bibitem[{\citenamefont{Migdal}(1962)}]{migdal}
\bibinfo{author}{\bibfnamefont{A.~B.} \bibnamefont{Migdal}},
  \emph{\bibinfo{title}{Theory of Finite Fermi Systems and Applications to
  Atomic Nuclei}} (\bibinfo{publisher}{Interscience}, \bibinfo{year}{1962}).

\bibitem[{\citenamefont{Baym and Pethick}(1991)}]{baym}
\bibinfo{author}{\bibfnamefont{G.}~\bibnamefont{Baym}} \bibnamefont{and}
  \bibinfo{author}{\bibfnamefont{C.~J.} \bibnamefont{Pethick}},
  \emph{\bibinfo{title}{Landau Fermi-Liquid Theory: Concepts and Applications}}
  (\bibinfo{publisher}{John Wiley \& Sons, INC.}, \bibinfo{address}{New York},
  \bibinfo{year}{1991}).

\bibitem[{\citenamefont{H.Heiselberg}(1993)}]{chris:93}
\bibinfo{author}{\bibfnamefont{C.~.~D.} \bibnamefont{H.Heiselberg}},
  \bibinfo{journal}{Ann. Phys..} \textbf{\bibinfo{volume}{223}},
  \bibinfo{pages}{37} (\bibinfo{year}{1993}).

\bibitem[{\citenamefont{Weinberg}(1967)}]{weinb}
\bibinfo{author}{\bibfnamefont{S.}~\bibnamefont{Weinberg}},
  \bibinfo{journal}{Phys. Rev. Lett.} \textbf{\bibinfo{volume}{19}},
  \bibinfo{pages}{1264} (\bibinfo{year}{1967}).

\bibitem[{\citenamefont{Salam}(1968)}]{salam}
\bibinfo{author}{\bibfnamefont{A.}~\bibnamefont{Salam}},
  \emph{\bibinfo{title}{Elementary particle Theory: Relativistic Groups and
  Analyticity (Nobel Symposium N8)}} (\bibinfo{publisher}{Almqvist and
  Wiksell}, \bibinfo{address}{Stockholm}, \bibinfo{year}{1968}).

\bibitem[{\citenamefont{Glashow}(1961)}]{glashow}
\bibinfo{author}{\bibfnamefont{S.~L.} \bibnamefont{Glashow}},
  \bibinfo{journal}{Nucl. Phys.} \textbf{\bibinfo{volume}{22}},
  \bibinfo{pages}{579} (\bibinfo{year}{1961}).

\bibitem[{\citenamefont{Bentz et~al.}(1990)\citenamefont{Bentz, Arima, and
  Baier}}]{arima1}
\bibinfo{author}{\bibfnamefont{W.}~\bibnamefont{Bentz}},
  \bibinfo{author}{\bibfnamefont{A.}~\bibnamefont{Arima}}, \bibnamefont{and}
  \bibinfo{author}{\bibfnamefont{H.}~\bibnamefont{Baier}},
  \bibinfo{journal}{Annals Phys.} \textbf{\bibinfo{volume}{200}},
  \bibinfo{pages}{127} (\bibinfo{year}{1990}).

\bibitem[{\citenamefont{Marcucci et~al.}(2001)}]{sciavila1}
\bibinfo{author}{\bibfnamefont{L.~E.} \bibnamefont{Marcucci}}
  \bibnamefont{et~al.}, \bibinfo{journal}{Phys. Rev.}
  \textbf{\bibinfo{volume}{C63}}, \bibinfo{pages}{015801}
  (\bibinfo{year}{2001}).

\bibitem[{\citenamefont{Lykasov et~al.}(2005)\citenamefont{Lykasov, Olsson, and
  Pethick}}]{chris2}
\bibinfo{author}{\bibfnamefont{G.~I.} \bibnamefont{Lykasov}},
  \bibinfo{author}{\bibfnamefont{E.}~\bibnamefont{Olsson}}, \bibnamefont{and}
  \bibinfo{author}{\bibfnamefont{C.~J.} \bibnamefont{Pethick}},
  \bibinfo{journal}{Phys. Rev.} \textbf{\bibinfo{volume}{C72}},
  \bibinfo{pages}{025805} (\bibinfo{year}{2005}).

\bibitem[{\citenamefont{Pines and Nozi\'{e}res}(1966)}]{pines}
\bibinfo{author}{\bibfnamefont{D.}~\bibnamefont{Pines}} \bibnamefont{and}
  \bibinfo{author}{\bibfnamefont{P.}~\bibnamefont{Nozi\'{e}res}},
  \emph{\bibinfo{title}{The Theory of Quantum Liquids, v.1}}
  (\bibinfo{publisher}{W.~A.~Benjamin, INC}, \bibinfo{year}{1966}).

\bibitem[{\citenamefont{B\'{a}ckman et~al.}(1979)\citenamefont{B\'{a}ckman,
  Sj\'{o}berg, and Jackson}}]{backman}
\bibinfo{author}{\bibfnamefont{S.~O.} \bibnamefont{B\'{a}ckman}},
  \bibinfo{author}{\bibfnamefont{O.}~\bibnamefont{Sj\'{o}berg}},
  \bibnamefont{and} \bibinfo{author}{\bibfnamefont{A.~D.}
  \bibnamefont{Jackson}}, \bibinfo{journal}{Nucl. Phys.}
  \textbf{\bibinfo{volume}{A321}}, \bibinfo{pages}{10} (\bibinfo{year}{1979}).

\bibitem[{\citenamefont{Schwenk et~al.}(2002)\citenamefont{Schwenk, Brown, and
  Friman}}]{brown:2003}
\bibinfo{author}{\bibfnamefont{A.}~\bibnamefont{Schwenk}},
  \bibinfo{author}{\bibfnamefont{G.~E.} \bibnamefont{Brown}}, \bibnamefont{and}
  \bibinfo{author}{\bibfnamefont{B.}~\bibnamefont{Friman}},
  \bibinfo{journal}{Nucl. Phys.} \textbf{\bibinfo{volume}{A703}},
  \bibinfo{pages}{745} (\bibinfo{year}{2002}).

\bibitem[{\citenamefont{Blaizot et~al.}(1976)\citenamefont{Blaizot, Gogny, and
  Grammaticos}}]{blaizot}
\bibinfo{author}{\bibfnamefont{J.~P.} \bibnamefont{Blaizot}},
  \bibinfo{author}{\bibfnamefont{D.}~\bibnamefont{Gogny}}, \bibnamefont{and}
  \bibinfo{author}{\bibfnamefont{B.}~\bibnamefont{Grammaticos}},
  \bibinfo{journal}{Nucl. Phys.} \textbf{\bibinfo{volume}{A265}},
  \bibinfo{pages}{315} (\bibinfo{year}{1976}).

\bibitem[{\citenamefont{Sj\'{o}berg}(1973)}]{sjoberg:1973}
\bibinfo{author}{\bibfnamefont{O.}~\bibnamefont{Sj\'{o}berg}},
  \bibinfo{journal}{Ann. \ Phys.} \textbf{\bibinfo{volume}{78}},
  \bibinfo{pages}{39} (\bibinfo{year}{1973}).

\bibitem[{\citenamefont{Sarsa et~al.}(2003)\citenamefont{Sarsa, Fantoni,
  Schmidt, and Pederiva}}]{fantoni:03}
\bibinfo{author}{\bibfnamefont{A.}~\bibnamefont{Sarsa}},
  \bibinfo{author}{\bibfnamefont{S.}~\bibnamefont{Fantoni}},
  \bibinfo{author}{\bibfnamefont{K.~E.} \bibnamefont{Schmidt}},
  \bibnamefont{and} \bibinfo{author}{\bibfnamefont{F.}~\bibnamefont{Pederiva}},
  \bibinfo{journal}{Phys. Rev.} \textbf{\bibinfo{volume}{C68}},
  \bibinfo{pages}{024308} (\bibinfo{year}{2003}).

\bibitem[{\citenamefont{Bahcall}()}]{Bahcall:2004fg}
\bibinfo{author}{\bibfnamefont{J.~N.} \bibnamefont{Bahcall}},
  \eprint{physics/0411190}.

\end{thebibliography}



\begin{figure*}[htb]
  \includegraphics[angle=270,width=0.80\linewidth]{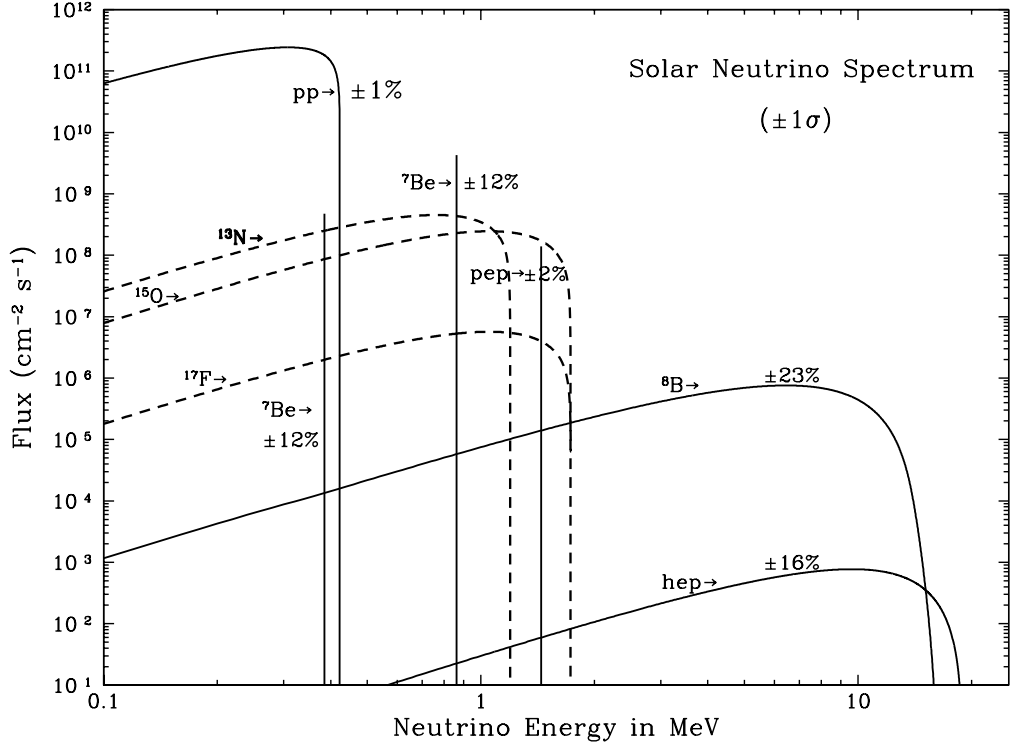}
\caption{\label{fig:fig3}
The illustration of the energy dependence of the solar neutrino
flux continuum from Ref.~\cite{Bahcall:2004fg}
}
\end{figure*}

\begin{figure*}[htb]
  \includegraphics[angle=270,width=0.80\linewidth]{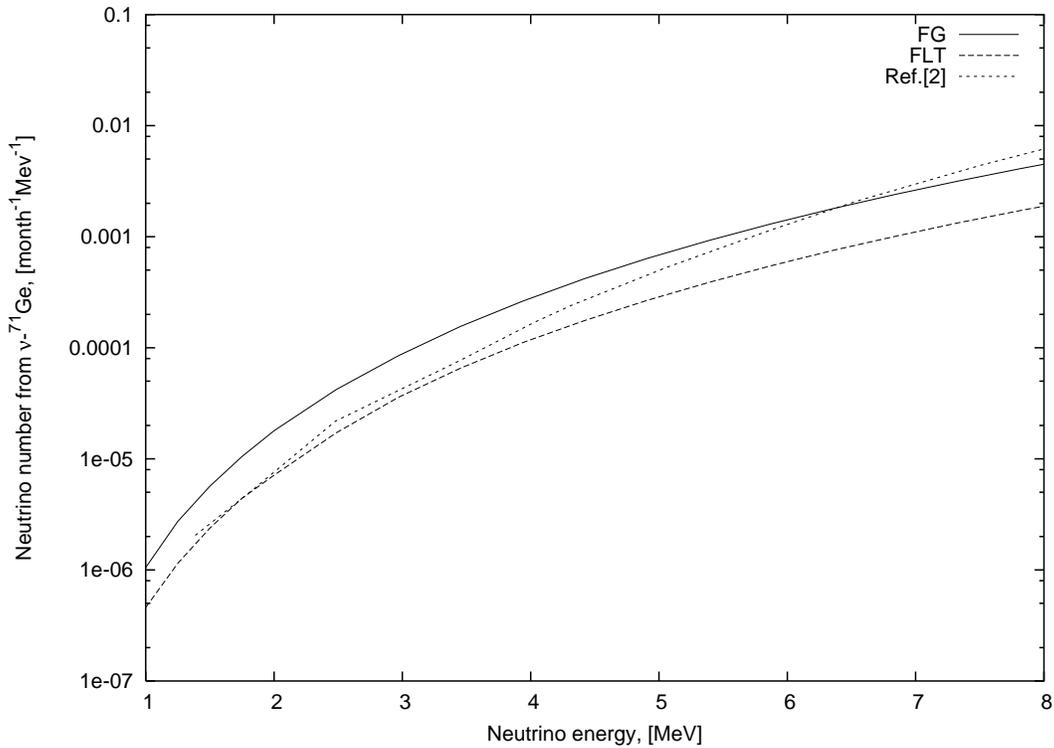}
\caption{\label{fig:fig4}
The total number of  neutrino interactions per month and per MeV produced from 
the $^8$B-$\nu$ flux  interacting with 1-kg $^{71}$Ge detector 
as a function of the neutrino energy $E_\nu$. Notation is the same as in Fig.~\ref{fig:fig1}
}
\end{figure*}

\end{document}